# Mapping the backbone of the Humanities through the eyes of Wikipedia

**Daniel Torres-Salinas**[a] E-mail: torressalinas@gmail.com

**Esteban Romero-Frías**[a] E-mail: erf@ugr.es

**Wenceslao Arroyo-Machado**[a] **[Corresponding author]** E-mail: wences@ugr.es

[a]Medialab UGR, University of Granada, Gran Vía de Colón, 48, 18071 Granada, Spain.

**Abstract**: The present study aims to establish a valid method by which to apply the theory of co-citations to Wikipedia article references and, subsequently, to map these relationships between scientific papers. This theory, originally applied to scientific literature, will be transferred to the digital environment of collective knowledge generation. To this end, a dataset containing Wikipedia references collected from Altmetric and Scopus' Journal Metrics journals has been used. The articles have been categorized according to the disciplines and specialties established in the All Science Journal Classification (ASJC). They have also been grouped by journal of publication. A set of articles in the Humanities, comprising 25 555 Wikipedia articles with 41 655 references to 32 245 resources, has been selected. Finally, a descriptive statistical study has been conducted and co-citations have been mapped using networks and indicators of degree and betweenness centrality.

## 1. Introduction

When Wikipedia was created in 2001 (DiBona, Cooper & Stone, 2006), few could have imagined that in a short time a voluntary, collective project would become the main encyclopedic work of reference for a large part of Humanity. The birth of Wikipedia, in the middle of the dot-com bubble, occurred during the prelude to the emergence of the Web 2.0 paradigm (O'Reilly, 2005) and was destined to become one of the greatest exponents of the Web's ability to activate the collective intelligence of Internet users (Surowiecki, 2005). In January 2018, 17 years later, the English language version of Wikipedia accounted for 5.5 million of the 46 million articles in the more than 170 editions of Wikipedia[1]. The Wikipedia in English—its largest edition—represents approximately 11.7% of the whole of Wikipedia, creating more than 600 new articles per day in 2017.

According to Alexa,[2] at the beginning of 2018, Wikipedia ranked 5th among the most visited websites in the world with a remarkable 66.4% of traffic received coming from user searches. These data refer to organic traffic received by the website and demonstrate that, for a wide variety of terms, Wikipedia is one of the first options that search engines offer as a relevant result on the Web. Hence, it constitutes a much-used reference resource that is of great importance for educational purposes in Science, the Humanities, and other fields. For example, as an encyclopedic digital project, Wikipedia is considered a "*very fertile ground for the creation of innovative projects related to the Digital Humanities*"[3]. It is argued that Wikipedia might be the best and the largest educational platform in history (Tramullas, 2016).

---

[1] https://en.wikipedia.org/wiki/Wikipedia:Size_of_Wikipedia (consulted on November 6, 2017)
[2] https://www.alexa.com/topsites (consulted on February 21, 2018)
[3] https://blog.wikimedia.org/2016/08/17/wikipedia-largest-digital-humanities-project/ (consulted on February 21, 2018)

Wikipedia is conceived of as a tool for the dissemination of knowledge through articles generated by its users under Creative Commons licenses (attribution-share alike). Wikipedia has overtaken its competitors by revolutionizing the industry through a profound epistemological transformation that focuses on the social dimension (Fallis, 2008, Fuchs, 2008). Over time, Wikipedia has developed complex rules—generated by the community itself—that are not rigid and remain subject to revision but, at the same time, are strictly observed. Articles should always be verifiable and have reliable sources. Insofar as encyclopedic content is concerned, secondary sources that are "*reliable, independent and published*" prevail. Among these, particular mention is made of specialized publications:

"*Many Wikipedia articles rely on scholarly material. When available, academic and peer-reviewed publications, scholarly monographs, and textbooks are usually the most reliable sources. However, some scholarly material may be outdated, in competition with alternative theories, or controversial within the relevant field. Try to cite current scholarly consensus when available, recognizing that this is often absent. Reliable non-academic sources may also be used in articles about scholarly issues, particularly material from high-quality mainstream publications. Deciding which sources are appropriate depends on context. Material should be attributed in-text where sources disagree.*"[4]

At the same time, from the perspective of scientific knowledge evaluation, in recent years digital indicators have been used as an alternative measure of academic impact: the so-called altmetrics indicators (Piwowar, 2013a, 2013b; Priem et al., 2010; Torres-Salinas, Cabezas-Clavijo & Jiménez-Contreras, 2013).

In this context, Wikipedia faces a dual challenge: on the one hand, the call to guarantee rigor in Wikipedia contents by referencing articles published in scientific journals; on the other, the opportunity to use Wikipedia references to scientific articles as a highly valuable altmetric information source to assess the social impact of research. Evidence of the value of references included in Wikipedia is its high weighting in a synthetic indicator such as the Altmetric Attention Score[5]. In this indicator, Wikipedia articles receive a rating of 3, which is higher than those corresponding to mentions on Twitter (1) or Facebook (0.25), but lower than references to news feeds (8) and blogs (5).

The connection between Wikipedia as a social platform and scientific articles has been explored in different ways. For example, through the analysis of reference and citation patterns in a specific scientific area (Serrano-López, Ingwersen & Sanz-Casado, 2017), as a platform for the promotion of open access scientific literature (Teplitskiy, Lu & Duede, 2016), or by exploring its limitations as a source in the evaluation of scientific activity (Kousha & Thelwall, 2016). Knowledge representation has also been formulated through reference maps connecting articles (Silva et al., 2011), or by analyzing differences between the Universal Decimal Classification (UDC) category structure and that generated by Wikipedia itself (Salah et al., 2012).

From a bibliometric perspective, co-citations constitute a classic instrument (Small, 1973) that allows knowledge to be mapped by taking account of common references received from a third document. Co-citations can be interpreted as a measure of the similarity between two documents. This approach has been used to observe the connections between words (Leydesdorff & Nerghes, 2017), or between areas of knowledge through scientific articles (Leydesdorff, Carley & Rafols, 2012). More recently, with the development of the Web, this concept has been transferred to this new space by discussing co-link analysis (Thelwall, 2009)—an approach based on sites or web pages that simultaneously link to other sites or web pages. Co-link analysis has proved a useful means of revealing the cognitive or intellectual structure of a field of study (Zuccala, 2006). Moreover, it has allowed investigators to broaden their scope of study beyond scientific production, having been applied to business (Vaughan & Romero-Frías, 2010), politics (Romero-Frías & Vaughan, 2010) or universities (Vaughan, Kipp & Gao, 2007).

---

[4] https://en.wikipedia.org/wiki/Wikipedia:Identifying_reliable_sources (consulted on February 21, 2018)

[5] https://help.altmetric.com/support/solutions/articles/6000060969-how-is-the-altmetric-attention-score-calculated-

In this regards, to our knowledge, no study has used Wikipedia as a reference to map science by extrapolating classical co-citation methodology to this digital platform in order to discover the structure of journals corresponding to different areas of knowledge and different scientific disciplines. With this approach, scientific knowledge could be mapped from a social perspective, thus offering a radically different view to that of the traditional maps constructed from the relationships between the scientific studies themselves. Based on this framework, we have focused on the Humanities in order to achieve the following objectives:

1. to establish a methodology to transfer co-citation theory to a digital environment taking as a reference an altmetric indicator linked to the collective generation of knowledge in Wikipedia;
2. to analyze how scientific knowledge is established in the field of the Humanities as this is represented in Wikipedia;
3. to evaluate the relationship between open access to articles in the Humanities and the citation of Humanities articles in Wikipedia.

## 2. Material and methods

### 2.1 Information sources and data processing

This study uses Altmetric.com as its source of information and the Altmetric Explorer to extract the references to scientific articles that are included in Wikipedia articles. To do this we have used the platform's download functions to obtain a csv file in which each scientific article appears with its basic data and information about the Wikipedia article in which it is referenced. So, all the scientific articles indexed in Altmetric.com and cited in Wikipedia have been downloaded. We have also used the Wikipedia API to obtain complementary information (ISSN).

A database with 261 079 Wikipedia entries was generated with a total of 1 214 322 references to 848 079 individual resources dated between 2004 and 2017. It should be noted that in 2004, 2005 and 2006 only 12 citations were counted. Only references for which Altmetric.com provides an associated publication date have been included, leading to an 8.8% (107 008) reduction in the dataset. When several citation dates were associated with the same Wikipedia article, only the most recent date has been taken into account, thus discarding duplications. Given the diversity of existing journals and their varied scientific nature, we decided to filter only those journals indexed in Scopus. Thus, we hoped to achieve two objectives: firstly, to guarantee that each reference corresponded to a valid scientific journal and, secondly, to obtain complementary information—such as the scientific category to which each article belonged. To do this, we used the Elsevier journal dataset in Cite Score Metrics,[6] indexed in 2016, as our source of information. Thus, the references were linked to the entire collection of Scopus journals. The final dataset contained 179 329 Wikipedia articles with 784 209 references to 549 782 individual resources, mainly scientific articles.

The present study focuses on scientific articles belonging to all 3209 journals in Scopus under the All Science Journal Classification (ASJC) code "Arts and Humanities" (discipline). Every journal within this discipline is attached to one or more specialties (subcodes within Scopus). Once our dataset had been merged with the Scopus data, our final sample comprised references to 1717 journals (54% of the total in Scopus), including: 25 555 articles (14.25% of all Wikipedia articles citing articles in Scopus included in all disciplines) with 41 655 references (5.31%) to 32 245 resources (5.86%). The vast majority (99.25%) of the articles in the final sample correspond to the English language Wikipedia; the rest are distributed between Swedish (0.6%) and Finnish (0.15%), taking no consideration of other languages. Figure 1 summarizes the process of collection and the evolution of sample size, as reported above.

---

[6] https://www.scopus.com/sources

Figure 1. Process of collection and the evolution of sample size

| | | | |
|---|---|---|---|
| Altmetric | 1 | Download resources cited in Wikipedia as indexed in Altmetric.com | 261 079 Wikipedia entries with 1 214 322 references |
| Altmetric | 2 | Download ISSN codes of resources by using the Altmetric.com API | |
| Scopus | 3 | Limitation of the sample to cited articles published in Scopus journals | 179 329 Wikipedia entries with 784 209 references to 549 782 articles |
| Scopus | 4 | Assignment of Scopus thematic category to entries and articles | |
| | 5 | Reduction of the sample to articles published in journals in the Humanities | 25 555 Wikipedia entries with 41 655 references to 32 245 articles |

Descriptive indicators have been calculated (mean, median, mode, standard deviation and range), as well as those corresponding to the degree centrality and betweenness centrality.

## 2.2 Development of science maps

In classic bibliometrics, the source of information is the scientific article. A co-citation is established when one scientific article cites another two articles, creating a relationship between them that could be interpreted as a measure of similarity between the authors, journals or the categories to which they belong (McCain, 1990). In the present study, we have used the co-citations established by Wikipedia entries (Figure 2) to allow us to draw a map of co-cited journals. Of the 1717 journals represented in the sample, 1408 were co-cited in the 13 specialties in the Humanities included in the Scopus classification.

For our analysis of the journals and specialties, we pruned these data by eliminating relationships with fewer than 6 co-citations in order to facilitate their visualization and interpretation. The vertices isolated in this process were subsequently eliminated. Finally, co-citation values were normalized to range between 0 and 1.

Figure 2. Diagram of co-citations from entries in Wikipedia

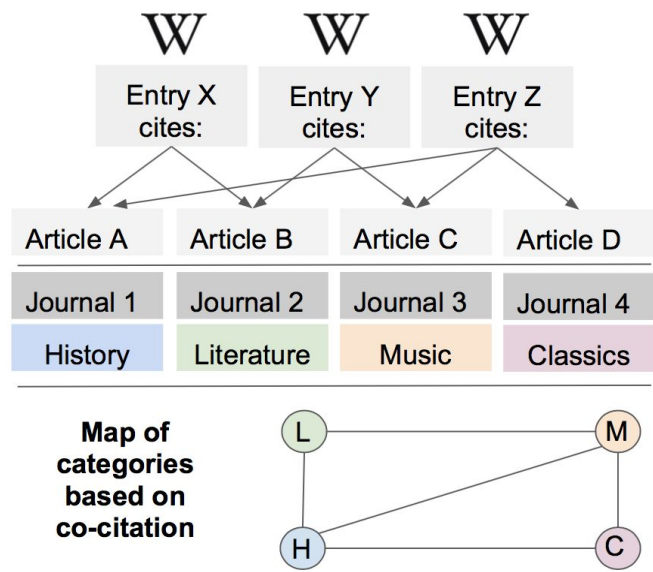

Next, we divided the data by components in order to extract the largest subset, consisting of 163 vertices. Finally, for both journals and specialties, we applied the Pathfinder algorithm (Vargas-Quesada, 2005), to identify the most relevant relationships based on triangular inequality, obtaining Pathfinder networks (PFNETs) with them. This technique has previously been used to map thematic domains in science (Moya-Anegón et al., 2004). As a result, two maps show how journals and specialties in the Humanities are linked to each other from the social perspective provided by Wikipedia. The tools used throughout this process were: Notepad ++, to correct and prepare the data downloaded from Altmetric.com through regular expressions; Microsoft Access, to store and treat data and for information retrieval; Microsoft Excel, for descriptive statistical analysis; Pajek, to elaborate maps and conduct the centrality study; Gephi, to design the maps; and the programming language R, to download data from the API and for data processing (for example, to combine categories using colors).

## 3. Analysis and results

### 3.1 General data and annual evolution

Table 1 shows descriptive statistics of the Wikipedia article references to scientific articles published in Scopus journals, and of the citations received by these scientific articles both for the whole of Wikipedia (global) and for the Humanities discipline. Note that we only take account of Wikipedia articles that include at least one citation to a scientific journal and scientific journals referenced at least once in Wikipedia. Hence, the minimum mean for references is 1. In total, 784 209 citations to scientific articles in all disciplines have been identified; of these 41 655 citations (5%) correspond to works in the Humanities. More specifically, 25 555 individual Wikipedia entries have been compiled, citing 32 245 independent articles. If we focus on the citations of scientific articles found in Wikipedia entries, we find a considerable difference between the global average for all disciplines (4.37) and that for the Humanities (1.63). In addition, there is greater homogeneity in terms of the average number of citations that articles receive: 1.42, globally, versus 1.29, for the Humanities.

Table 1. Descriptive statistical analysis of the distribution of references and citations in Wikipedia articles globally and for the Humanities

| **References to scientific articles included in Wikipedia entries*** | **Global** | **Humanities** |
|---|---:|---:|
| Mean | 4.37 | 1.63 |
| Median | 2 | 1 |
| Standard deviation | 8.25 | 1.76 |
| Range | 440 | 54 |
| Total entries with at least 1 reference | 179 329 | 25 555 |
| Total citations in Wikipedia | 784 209 | 41 655 |
| | | |
| **Citations of scientific articles received from Wikipedia*** | **Global** | **Humanities** |
| Mean | 1.42 | 1.29 |
| Median | 1 | 1 |
| Standard deviation | 10.59 | 1.23 |
| Range | 5067 | 106 |
| Total articles cited | 549 782 | 32 245 |

If we depict the annual evolution of the Humanities, the number of citations has been especially dense since 2014: the period 2007-2013 saw some 2500 citations annually; however, since 2014, this has increased to around 7500 citations per year. The most active year was 2016 with 8464 citations. The average number of citations received per scientific article has also shown a positive growth trend, reaching its highest level in 2016 (mean 1.37) and 2017 (mean 1.37).

Figure 3. Annual evolution of the number of citations included in Wikipedia and the average number of citations received per article in the Humanities during the period 2007-2017

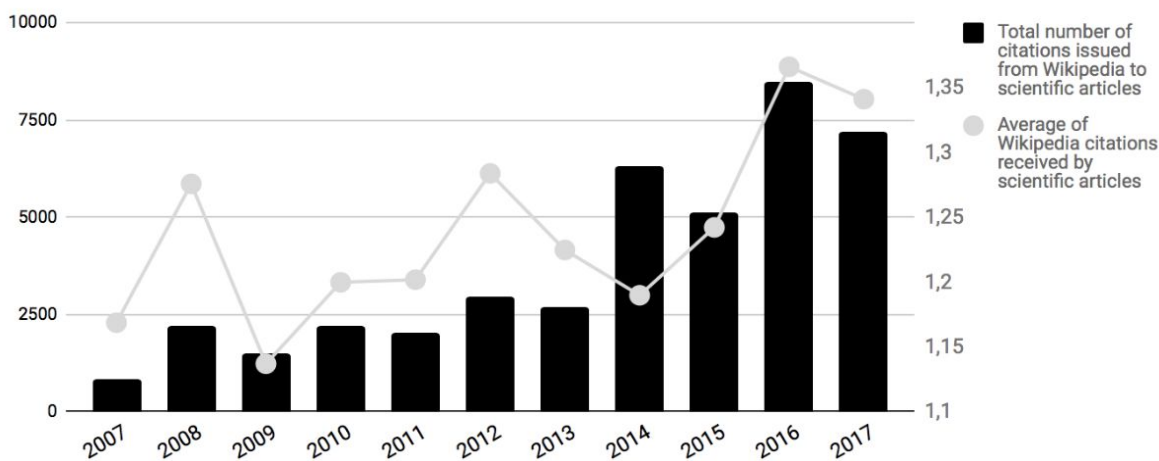

## 3.2 Analysis of specialties in the Humanities

Table 2 shows the Scopus classification specialties in the Humanities, allowing us to identify those that receive most attention in Wikipedia. The most outstanding, at a considerable distance from the rest, is *History*, which concentrates the largest number of single journals cited (531), scientific articles cited (11 661) and total citations (15 969). The specialties *Language and Linguistics* and *History & Philosophy of Science* stand out in terms of the number of citations (without considering the miscellaneous category *Arts & Humanities*). The *Museology* category, despite being smaller, receives higher average citations per article (1.43). However, the average number of citations per article is generally quite homogeneous, ranging between 1.18, corresponding to *Literature and Literary Theory*, and the aforementioned 1.43, corresponding to *Museology*. Among the specialties that receive less attention are *Classics* and *Conservation*, which account for only 1.4% and 0.4%, respectively, of the total number of citations in Wikipedia.

Table 2. Citation indicators of journals and articles referenced in Wikipedia for specialties in the Humanities

|  | No. of journals cited in Wikipedia indexed in Scopus | | No. scientific articles cited in Wikipedia | Total number of citations received in Wikipedia | | Average number of citations in Wikipedia received by article |
|---|---|---|---|---|---|---|
| Archeology | 111 | 4.9% | 2026 | 2785 | 5.2% | 1.37 ± 1.03 |
| Arts and Humanities | 266 | 19% | 7881 | 10 034 | 18.7% | 1.27 ± 0.86 |
| Classics | 37 | 1.4% | 589 | 744 | 1.4% | 1.26 ± 0.85 |
| Conservation | 18 | 0.3% | 144 | 188 | 0.4% | 1.30 ± 1.42 |
| History | 531 | 28.1% | 1661 | 15 969 | 29.8% | 1.36 ± 1.76 |
| History and Philosophy of Science | 90 | 8.5% | 3524 | 4574 | 8.5% | 1.29 ± 0.88 |
| Language and Linguistics | 261 | 9.6% | 3990 | 4796 | 9% | 1.20 ± 0.65 |
| Literature and Literary Theory | 282 | 7.5% | 3140 | 3729 | 7% | 1.18 ± 0.68 |
| Museology | 17 | 1.9% | 811 | 1166 | 2.2% | 1.43 ± 2.03 |
| Music | 65 | 2.8% | 1194 | 1473 | 2.8% | 1.23 ± 0.75 |
| Philosophy | 211 | 6.2% | 2588 | 3147 | 5.9% | 1.21 ± 0.66 |
| Religious studies | 170 | 4.2% | 1723 | 2190 | 4.1% | 1.27 ± 0.87 |
| Visual Arts and Performing Arts | 188 | 5.3% | 2197 | 2719 | 5.1% | 1.23 ± 1.32 |

Figure 4 shows the co-citation map for specialties in the Humanities after editing the data following the application of the Pathfinder algorithm. In this map, the thickness of the edges indicates the degree of co-citation. The size of the nodes represents the number of articles within the specialty that establish a co-citation. Note the scarce connection between specialties with highly homogeneous citation patterns, in regards to specialties, in Wikipedia entries. *History* occupies a highly relevant position as it is related to 11 specialties, showing the strongest links with the categories of *History and Literary Theory*, *History and Philosophy of Science* and the miscellaneous *Arts and Humanities*. The only two specialties not linked to *History* are *Music* and *Language and Linguistics*, directly connected to *Arts and Humanities*.

Figure 4. Co-citation map of specialties in the Humanities from the co-citations received from Wikipedia entries during the period 2007-2017 using the Pathfinder algorithm

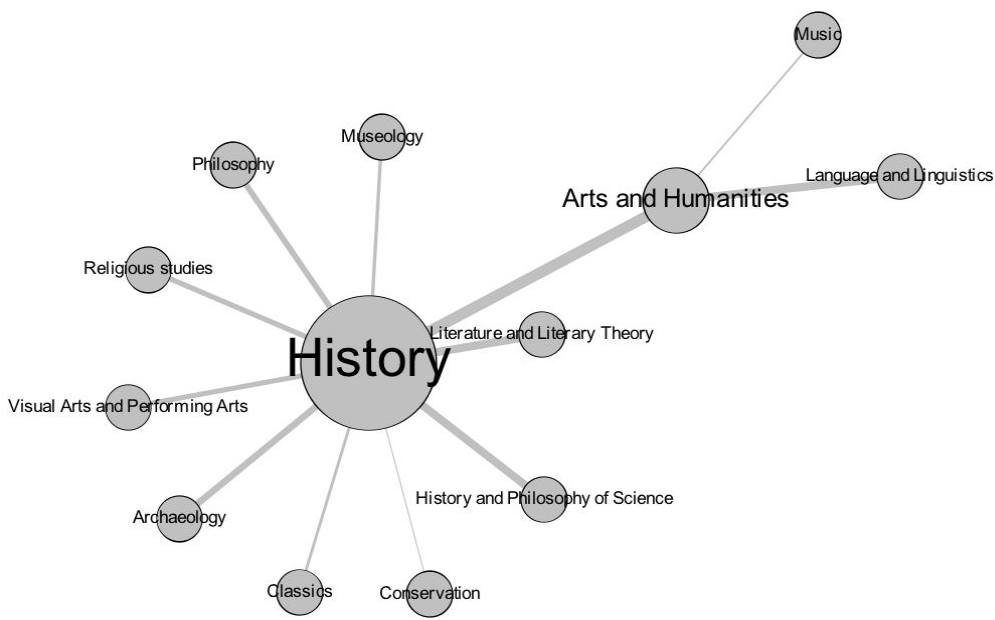

### 3.3 Analysis of journals in the Humanities

Table 3 lists the first 25 journals ordered according to the number of citations received from Wikipedia. We would conclude that these are among the publications with higher social use on this platform. The journal that receives the highest number of citations (869) is *Annals of the New York Academy of Sciences*. This is a multidisciplinary journal, founded in 1823, that publishes on biomedicine and biology, but also on philosophy and anthropology. The profile for the remaining journals is not homogeneous, including topics like: History (*English Historical Review, American Historical Review*), Anthropology (*Current Anthropology*), Linguistics (*International Journal of American Linguistics*) or multidisciplinary topics such as sex (*Archives of Sexual Behavior*). Note that none of these journals is published in open access, in marked contrast to the open nature of the encyclopedia. It is also remarkable that 18 of these publications are high impact journals because they are among the top 10 of those with the greatest impact in their specialty according to the Scopus Journal of Metrics. Therefore, we can conclude that Wikipedia editors consider that journals with higher impact on the scientific community are also more reliable sources of information.

Table 3. Most cited journals in the Humanities in Wikipedia during the period 2007-2017

| | | No. of citations received in Wikipedia | No. of articles cited in Wikipedia | Average number of citations per article | Open Access journal? | Top journal?** |
|---|---|---|---|---|---|---|
| 1 | Annals of the New York Academy of Sciences | 869 | 698 | 1.24 | No | Yes |
| 2 | Journal of the Acoustical Society of America | 621 | 519 | 1.20 | No | No |
| 3 | Archives of Sexual Behavior | 591 | 373 | 1.58 | No | No |
| 4 | Isis | 502 | 357 | 1.41 | No | Yes |
| 5 | English Historical Review | 499 | 320 | 1.56 | No | Yes |
| 6 | American Historical Review | 444 | 378 | 1.17 | No | Yes |
| 7 | Current Anthropology | 416 | 271 | 1.54 | No | Yes |
| 8 | Journal of Archaeological Science | 396 | 270 | 1.47 | No | Yes |
| 9 | Quaternary Science Reviews | 355 | 260 | 1.37 | No | Yes |
| 10 | Social Science and Medicine | 333 | 267 | 1.25 | No | Yes |
| 11 | American Museum Novitates | 333 | 128 | 2.60 | No | Yes |
| 12 | Journal of the American Oriental Society | 322 | 215 | 1.50 | No | No |
| 13 | Bulletin of the School of Oriental and African Studies | 316 | 219 | 1.44 | No | No |
| 14 | Cognition | 296 | 221 | 1.34 | No | Yes |
| 15 | Intelligence | 291 | 203 | 1.43 | No | Yes |
| 16 | Speculum | 287 | 215 | 1.33 | No | Yes |
| 17 | Journal of Asian Studies | 275 | 196 | 1.40 | No | Yes |
| 18 | International Journal of American Linguistics | 271 | 218 | 1.24 | No | No |
| 19 | Medical History | 270 | 187 | 1.44 | No | Yes |
| 20 | Language | 255 | 190 | 1.34 | No | No |
| 21 | Economic History Review | 243 | 117 | 02.08 | No | Yes |
| 22 | Journal of American History | 227 | 191 | 1.19 | No | No |
| 23 | American Antiquity | 224 | 164 | 1.37 | No | Yes |
| 24 | Journal of Sex Research | 218 | 144 | 1.51 | No | Yes |
| 25 | Journal of African History | 215 | 147 | 1.46 | No | Yes |
| ** Top is defined as being among the 10% most cited journals in the Scopus/Elsevier Score Metrics categories | | | | | | |

Figure 5 shows the co-citation map between journals. The scientific journals in the Humanities cited in Wikipedia have been grouped into 10 clusters each of which is represented by a color. If we first consider the specialties, not all the clusters are homogeneous as they are composed of journals from different specialties. However we should distinguish between clusters with a lower degree of heterogeneity (0, 1, 3 or 4) and more heterogeneous clusters (5, 8 or 9). Clusters 6 and 9 are identified at the center of the network with a mediating role and connecting specialties. In these two clusters we find multidisciplinary journals belonging mainly to three areas, *History*, *Archeology* and *Linguistics*. Cluster 9 connects with cluster 0, which includes journals from *Language and Linguistics*, and cluster 1, which includes *Philosophy of Science*. Cluster 6 connects with clusters 5 and 2—formed by *History* and *Philosophy of Science*—and clusters 7 and 8—also formed in the main by history journals. To summarize, in this representation of knowledge from Wikipedia, the upper part (Clusters 0, 9, 1 and 3) represents *Language and Linguistics* and *History and Philosophy of Science*; the lower part is dominated by *History* and *Archeology*, although it is closely related to other specialties.

Figure 5. Map of co-citation in Wikipedia of scientific journals in the Humanities grouped according to similarity clusters

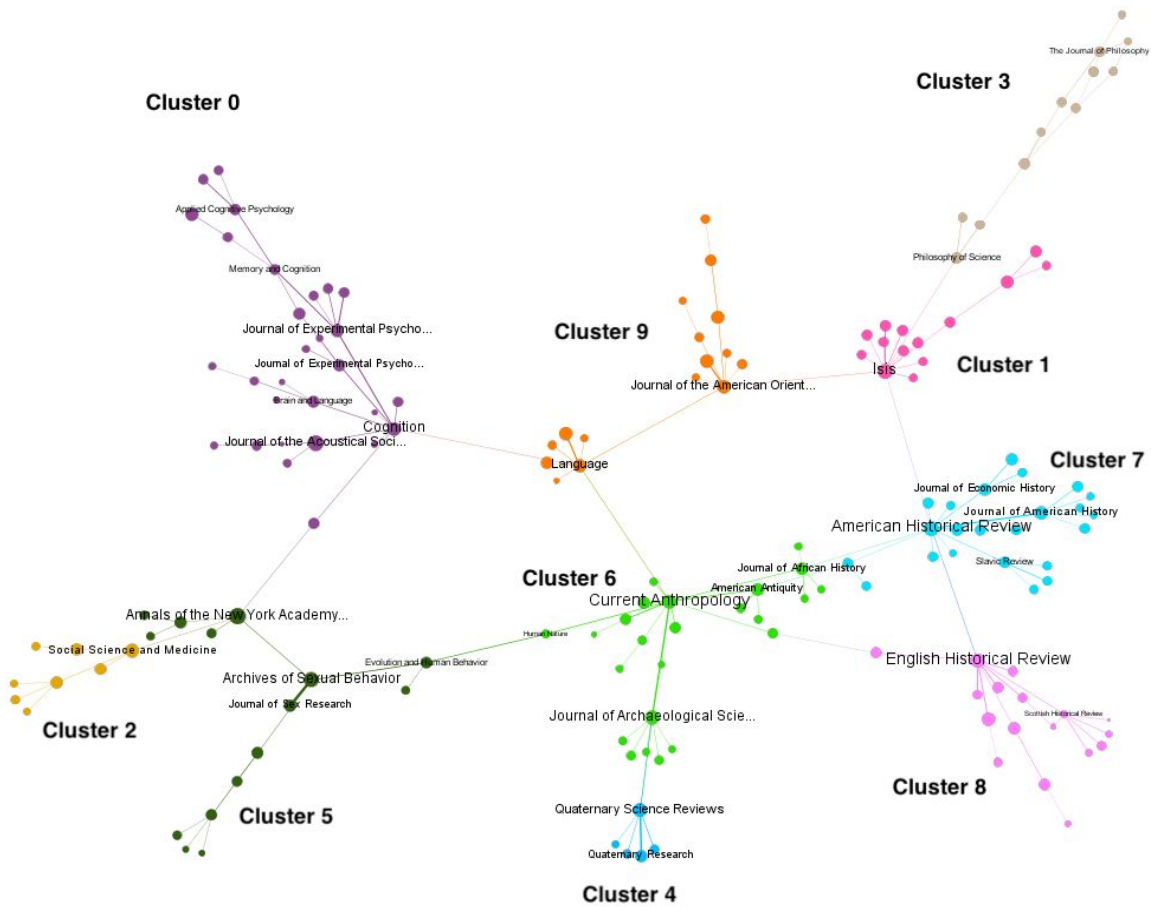

| **Cluster 0** | **Cluster 1** | **Cluster 2** | **Cluster 3** | **Cluster 4** |
|---|---|---|---|---|
| Arts & Humanities: 59% Language & Linguistics: 36% | History: 46% History & Phil. Science: 46% | Arts & Humanities: 62% History: 12% History & Phil. Science: 12% Philosophy: 12% | Philosophy: 80% History & Phil. Science: 14% | Archaeology: 60% Arts & Humanities: 40% |
| **Cluster 5** | **Cluster 6** | **Cluster 7** | **Cluster 8** | **Cluster 9** |
| Arts & Humanities: 73% History & Phil. Science: 13% History: 6.7% Literature & Lit. Theory: 6.7% | Archaeology: 35% History: 32% Arts & Humanities: 31% | History: 75% Arts & Humanities: 9% Literature & Lit. Theory: 7.27 | History: 73% Archaeology: 10% Language & Linguistics: 6% Religious studies: 4% | Language & Linguistics 37% History 28% Arts & Humanities 12% Religious studies 12% Philosophy 6% |

### 3.4 Comparison with other studies

Following we compare our results with those from similar studies. For instance, Richardson (2013) used the same database and thematic categorization. To compare our results with his we have replicated the same methodology, but applying it to the Wikipedia data with the aim of getting a similar map. In this way, we have generated a co-citation network of journals in Wikipedia entries, it is composed of 1408 nodes and 12 131 edges, which has been filtered to its main component of 1388 nodes and 12 121 edges. Figure 6 shows how the position of *History* is kept with a weight and role more determinant than the rest of specialties. *History* is positioned in the center of the network and is highly connected to other specialties (*Archeology* and *History and Philosophy of Science*). It is important to highlight the secondary role of *Literature and Literary Theory*, with a much smaller size, which is relegated to the periphery of the network and loosely connected. The main role of *History* and the secondary role of *Literature and Literary Theory* are the principal differences found in relation with the studies of Richardson (2013) and Leydesdorff et al. (2011).

Figure 6. Main component of the network of co-citation humanities journals in Wikipedia

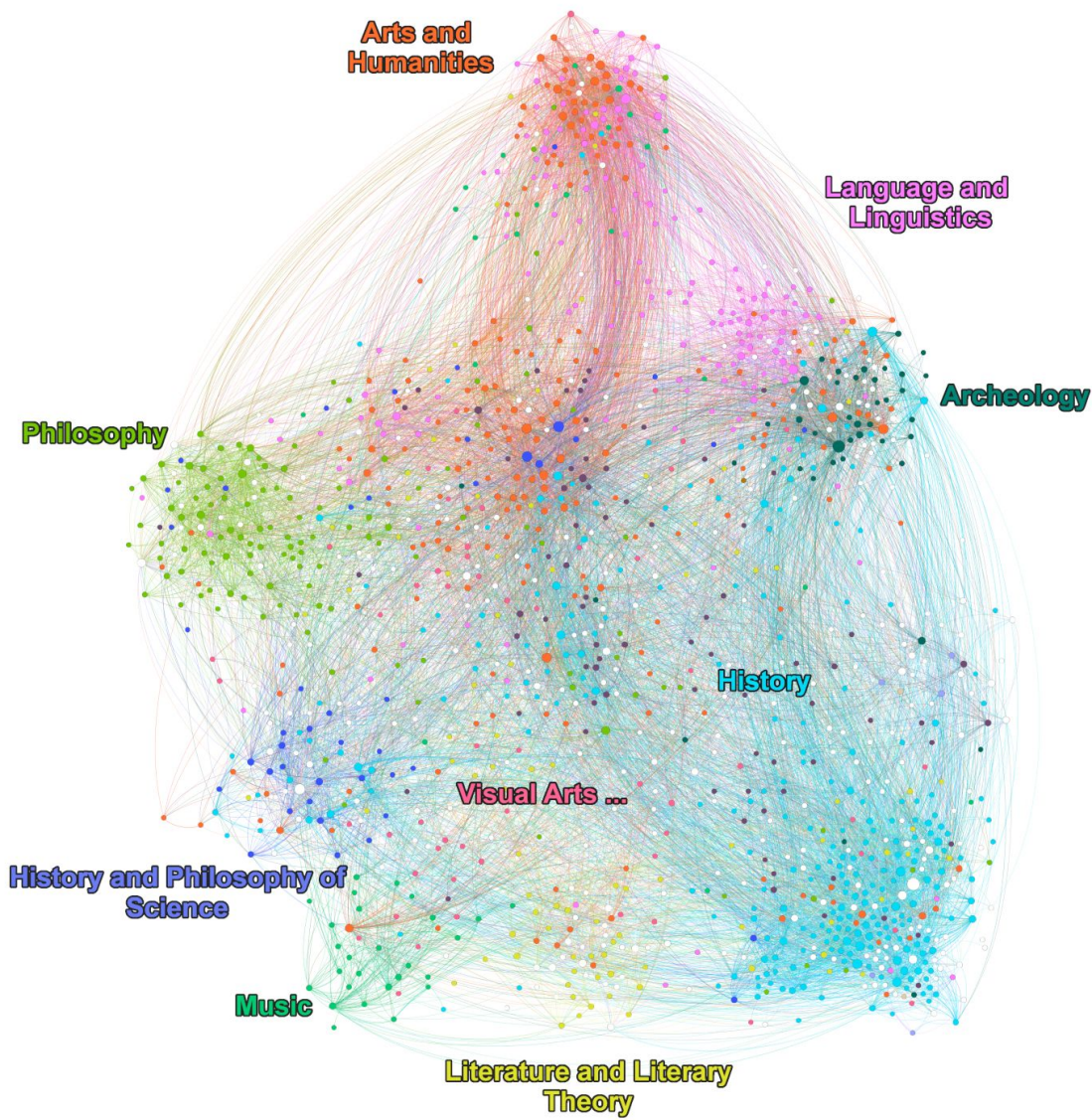

Colours correspond to journals specialties, the journals with more of one area are white. Due to the low presence of the journals with the unique area of *Classics*, *Conservation* and *Museology*, they aren't tagged in the network, while *Religious studies* are too disseminated for it.

Hence, there are evident differences between the two studies. Following we provide a plausible explanation for them. The maps generated are different due to the different coverage size and proportion of Scopus and Wikipedia. As observed in table 4, History accumulates 28% of the total number of Wikipedia articles, while in Scopus, History represent only 11% of the database. This fact contrasts, for example, with the case of Literature and Literary Theory, which has 7.57% papers of Wikipedia while in Scopus represents 14.18% of the database. This affects the positioning and degree of these specialties in the two networks. Furthermore, it evidences that social interest does not always align with scientific interest.

Table 4. Coverage of humanities specialties in Wikipedia and Scopus in function to the number of articles and cites

|  | Coverage Wikipedia | | Coverage Scopus | |
| --- | --- | --- | --- | --- |
|  | % of articles | % of citations | % of articles | % of citations |
| History | 28.12% | 29.84% | 17.39% | 11.70% |
| Arts and Humanities | 19.01% | 18.75% | 15.41% | 35.31% |
| Language and Linguistics | 9.62% | 8.96% | 11.89% | 16.01% |
| History and Philosophy of Science | 8.5% | 8.55% | 3.82% | 8.20% |
| Literature and Literary Theory | 7.57% | 6.97% | 14.18% | 2.92% |
| Philosophy | 6.24% | 5.88% | 10.76% | 7.78% |
| Visual Arts and Performing Arts | 5.30% | 5.08% | 8.83% | 2.43% |
| Archaeology | 4.89% | 5.20% | 5.15% | 10.10% |
| Religious studies | 4.16% | 4.09% | 7.07% | 2.66% |
| Music | 2.88% | 2.75% | 2.28% | 1.17% |
| Museology | 1.96% | 2.18% | 0.72% | 0.45% |
| Classics | 1.42% | 1.39% | 1.32% | 0.26% |
| Conservation | 0.35% | 0.35% | 1.17% | 1.01% |

*Scopus citation data from the CiteScore 2016

## 4. Conclusions

In the present study, we have extrapolated the methodology for representing science on the basis of co-citation maps to a different context. Traditionally, science maps have been drawn up from scientific articles, using large databases such as the Web of Science or Scopus and demonstrating their validity as a means of establishing relationships between areas and of determining the structure of science from the scientific knowledge itself (Noyons & Van Raan, 1998). In the present study, these co-citation techniques have been extrapolated to a digital, social environment—Wikipedia—illustrating the use of articles as a source of citizen information, and the vision of the structure—from a social point of view—of scientific knowledge. More specifically, a vision of the Humanities has been shown from Wikipedia, the main encyclopedic project, based on collaborative and open principles.

The mapping technique has been successfully extrapolated to create co-citation maps based on categories pruned by applying the Pathfinder algorithm proposed by Moya-Anegon et al. (2004), showing that social platforms can be used to offer an alternative vision of scientific knowledge. However, it should be noted that the methodology used, which combines various sources (Altmetric.com, Wikipedia and Journal Metrics by Elsevier), has some limitations. For example, only scientific articles have been taken into account since only those resources with an ISSN and indexed by Scopus have been used, thus excluding books or chapters of special relevance in the Humanities (Torres-Salinas et al., 2013). This problem is present in other classical approaches that are limited to scientific journals (Leydesdorff, Hammarfelt & Salah, 2011).

Science maps based on categories pre-assigned by databases always offer a biased view since journals and studies do not always belong to the category assigned by the database (Rafols, Porter & Leydesdorff, 2010). An obvious example in the classification of the Scopus ASJC is the use of insignificant generic categories such as *Arts and Humanities (miscellaneous)* and *General Arts and Humanities*, which we had to unify under the label *Arts and Humanities*. Likewise, because journals may have more than one specialty assigned, the problem of latent co-citation arose (Vargas-Quesada, 2005), which we solved by combining all of them under the same label.

Despite its limitations, this study has served to illustrate the use of scientific information in a social context; for example, we have determined that the mean of works in the Humanities cited in Wikipedia is lower than the general mean including all the areas. Also, only 5% of the 784 209 citations in Wikipedia of scientific articles in Scopus correspond to articles in journals in the Humanities. This could suggest the need to strengthen the visibility of work in the Humanities so that it achieves greater social impact. It is well worth noting that since 2013 the annual evolution of citations in the Humanities has risen from an average of 2500 to 7500 per year. Also, despite the open philosophy of Wikipedia—a platform that works thanks to the legal support provided by Creative Commons licenses—the data indicate that of the 25 most cited journals on Wikipedia, none is open access, while more than 70% are among the 10% most cited in their category, with scientific quality prevailing in the open access model.

In relation to the maps, if we look at the specific categories within the Humanities, *History* is presented as the main specialty from a social point of view. It concentrates the largest number of citations of individual journals (531) and scientific articles (11 661), and the highest number of total citations (15 969). Co-citation analysis also places it in a central position, connecting specialties. Important connections between specialties have been determined, such as those between *History* and *Archeology* (Cluster 6) and *History* and *Language and Linguistics* (Cluster 9), around which the other specialties are articulated. *Philosophy* and *Philosophy of Science* are less well represented and occupy more peripheral positions than other specialties (for example Clusters 1, 2 and 3).

If we relate this social vision of science with more traditional bibliometric studies (Richardson, 2013; Leydesdorff, Hammarfelt & Salah, 2011), we encounter interesting differences. Richardson (2013), taking data from Scopus citations in 1570 journals in the *Arts & Humanities*, formulated a map in which the various themes are grouped around *Literature and Arts*, which occupies a central position. They are closely connected with *History*, a specialty that does not occupy as central a position as in our analysis.

On the other hand, Leydesdorff, Hammarfelt and Salah (2011) used Web of Science data to map relationships between 1157 *Arts & Humanities Citation Index* journals in 2008. They observed that *Literature* continued to occupy a central position connecting categories such as *Music*, *Philosophy*, *Linguistics*, *Art* and *History*. *History* in this study was subdivided into three parts: *American History*, *History and Philosophy of Science* and *History*, properly speaking. Although it was more centrally positioned than in Richardson's study (2013), it was far from the nuclear role it occupies in our research. This is an indicator of how, from a social point of view, History is the key specialty that connects with other areas of humanistic knowledge and may reflect how the consumption of information and its relationships can differ in a social context by comparison with a scientific context.

To conclude, firstly, a reproducible methodology has been proposed to map scientific knowledge in Wikipedia through bibliometric techniques while, secondly, we have been able to analyze how the "global brain" perceives scientific knowledge and the interrelationship between specialties, offering a new vision of science as a counterpoint to the traditional maps. This methodology, based on the combination of sources such as Altmetric and Scopus, opens the door to other analyses drawing on sources such as Twitter, the News (news feeds) or report (policy feeds) that reflect the social vision of science from different perspectives (social, political, the mass media, among others).

## Acknowlegements

This work has been possible thanks to financial support from "Knowmetrics: knowledge evaluation in digital society", a project funded by scientific research team grants from the BBVA Foundation, 2016. We thank Altmetric.com for the transfer of the data that has allowed us to conduct this study.